\documentclass[%
 aip,
 amsmath,amssymb,
 reprint,%
]{revtex4-1}
\usepackage{graphicx}
\usepackage{dcolumn}
\usepackage{bm}
\usepackage[normalem]{ulem}
\usepackage[utf8]{inputenc}
\usepackage[T1]{fontenc}
\usepackage{mathptmx}
\usepackage{etoolbox}
\usepackage{hyperref}
\usepackage{blindtext}
\usepackage{lipsum}
\usepackage{multirow} 
\usepackage{subcaption} 
\usepackage{slashbox}
\usepackage{placeins}
\usepackage{ragged2e}
\usepackage{graphicx,subcaption}
\hypersetup{colorlinks,citecolor=blue,filecolor=blue,linkcolor=blue,urlcolor=blue}

\makeatletter
\def\@email#1#2{%
 \endgroup
 \patchcmd{\titleblock@produce}
  {\frontmatter@RRAPformat}
  {\frontmatter@RRAPformat{\produce@RRAP{*#1\href{mailto:#2}{#2}}}\frontmatter@RRAPformat}
  {}{}
}%
\makeatother
\begin{document}


\title{Dipole Propagation in Inhomogeneous Strongly Coupled Dusty Plasmas: A Viscoelastic Fluid Approach }

\author{Vipul B. Rohit}
\affiliation{Department of Physics, Sardar Vallabhbhai National Institute of Technology, Surat-395007, Gujarat, India}
 \email{d22ph002@phy.svnit.ac.in}
\author{Vikram S. Dharodi}%
\affiliation{Department of Physics and Astronomy, West Virginia University, Morgantown, WV, 26506, USA}%
\email{vikram.ipr@gmail.com}
\author{Sharad K. Yadav}
\affiliation{Department of Physics, Sardar Vallabhbhai National Institute of Technology, Surat-395007, Gujarat, India}%
\email{sharadyadav@phy.svnit.ac.in}
\date{\today}

\begin{abstract}
The propagation characteristics of fluid vortices—particularly monopoles and dipoles—in a homogeneous viscoelastic fluid were reported in a recent publication [Dharodi {\it{et. al.}} Phys. Plasmas 23, 013707 (2016)]. In that study, a dusty plasma was modeled as a viscoelastic fluid using the incompressible limit of the generalized hydrodynamic model under strongly coupled conditions—i.e., in a regime where the system remains in a fluid state but exhibits significant interparticle correlations, with potential energy dominating over kinetic energy. In this paper, we extend the previous work by employing the same model to investigate the evolution of a dipole—represented by two counter-rotating Lamb-Oseen vortices—in an inhomogeneous medium. It is shown that the entire dynamics of a dipole is governed by the competition between the strength of transverse shear waves, which is proportional to the elastic strength of the viscoelastic background medium, and the circulation strength of the vortices in the dipole structure. The density inhomogeneity is introduced along the vertical y-direction, perpendicular to the direction of dipole motion, using both smooth and sharp cutoffs.  The numerical simulations show that a higher circulation strength of a dipole and/or lower coupling strength of the background medium allows the dipole to survive longer and follow a more pronounced curved trajectory toward the high-density side. While the overall effects of circulation and coupling strength are similar in both densities, the resulting structure morphologies differ: in the smooth density case, the interface around the vortices gradually forms a mushroom-like shape, whereas in the sharp case, it forms a simple spiral envelope. These effects are visualized through two dimensional simulations based on the incompressible generalized hydrodynamic model.
\end{abstract}

\maketitle
\section{Introduction}
Two counter-rotating vortices, collectively known as a propagating dipole structure, play a crucial role in enhancing advection and material transport as they traverse a medium. Typical examples where dipole vortices form and influence advection and transport include Rayleigh–Taylor instability, where density differences between fluid layers generate dipolar vortices \cite{weijermars2021diffusive}; Rayleigh–Bénard convection, where thermal plumes driven by temperature gradients produce dipole structures \cite{zhou2007morphological,pieri2016plume}; aerodynamic lift, where pressure differentials across a wing create dipole vortices in the wake \cite{devenport1997structure,chan2011vortex}; storm systems, where pressure gradients and Coriolis forces form vortex pairs \cite{rotunno1982influence,klemp1987dynamics}; turbulence, where complex nonlinear interactions between different scales spontaneously generate dipole vortices \cite{mcwilliams1984emergence}; and plasmas, where magnetic reconnection launches bi-directional plasma jets with dipolar characteristics due to the breaking and reconnection of magnetic field lines \cite{yamada2010magnetic}. Thus, the critical role of dipole vortices in transport and dynamics has made their study a focus of extensive research across various flow environments, including hydrodynamic fluids \cite{hopfinger1993vortices,van1989dipole,flor1994experimental,flaor1995decay,trieling1998dipolar,davies2023spontaneous}, geophysical flows \cite{fedorov1989mushroom,weiss2001coherent}, plasmas \cite{fontan1995dynamics}, and astrophysical systems \cite{davies2023deformation}. 
The primary objective of this study is to investigate the propagation of a dipolar vortex structure in a two-dimensional (2D) strongly coupled dusty plasma (SCDP) with inhomogeneous density. The motivation for choosing 2D dusty plasmas includes their preference in laboratory experiments and simulations and their ability to easily reach strong coupling states due to highly charged dust particles \cite{thomas1994plasma,thomas1999direct,ichiki2004melting,haralson2015laser,haralson2016temperature,bandyopadhyay2008visco,chaubey2022preservation,chaubey2022coulomb,singh2023experimental,singh2023confinement}. A dusty plasma with strong coupling acts like a viscoelastic (VE) fluid below the crystallization limit \cite{kaw1998low,bandyopadhyay2008visco}. The presence of elasticity reduces the damping impact of viscosity, resulting in a longer lifespan of coherent structures \cite{frenkel_kinetic}. So, in a dusty plasma medium with strong coupling, dipole structures can move for long periods of time with little loss compared to a simple viscous fluid. Coherent or Vortical structures, including dipole vortices in dusty plasmas, are persistent formations that can arise from waves and/or instabilities interactions \cite{dharodi2016collective,dharodi2022kelvin,kumar2023kelvin}, external fields \cite{law1998probe}, or imbalances in charge or density \cite{choudhary2025review}.
Extensive research has investigated dipolar vortices (counter-rotating vortices) in both dusty plasmas - using computational \cite{singh2014visco,dharodi2016sub}, experimental \cite{choudhary2017experimental,choudhary2020three}, and analytical \cite{bharuthram1992vortices, rawat1993kelvin, shukla1993vortex, jovanovic2001dipolar,shukla2003solitons,ijaz2007vortex,laishram2021driven} approaches - and in hydrodynamic (HD) fluids - through computational \cite{nielsen1997formation,schmidt1998interaction,beckers2002dipole}, experimental \cite{couder1986experimental,fuentes1994experimental}, and analytical \cite{swaters1988viscous} methods. In dusty plasmas: Bharuthram et al. analytically predicted dipolar vortices as stationary solutions of coupled ion velocity gradient fluctuations in magnetized dusty plasmas using a two-fluid model \cite{bharuthram1992vortices}. Shukla et al., using a kinetic description for the ion fluid and hydrodynamic models for the electron and dust fluids, found that stationary solutions of nonlinear equations are dipolar vortices in a magnetized dusty plasma \cite{shukla1993vortex}. In a recent dusty plasma experiment, Choudhary et al. observed a pair of counter-rotating vortices driven by the 
${\bf{E}}{\times}{\bf{B}}$ drift \cite{choudhary2020three}. In hydrodynamic fluids: Nielsen et al. \cite{nielsen1997formation} conducted numerical studies on the formation and evolution of dipolar vortex structures in two-dimensional flows governed by the Navier-Stokes equations. Their results showed that, depending on the initial conditions, the dipoles typically evolve into Lamb-dipole-like structures, and they also examined the role of viscosity in their decay. Beckers et al. \cite{beckers2002dipole} examined  the interaction between two counter-rotating vortices in a stratified fluid using numerical simulations and laboratory experiments, focusing on the influence of Reynolds and Froude numbers.

A SCDP as a VE fluid supports transverse shear (TS) and longitudinal modes \cite{kaw2001collective,kumar2019coupling,choudhary2020influence}. To isolate the effect of inhomogeneity on transverse modes and avoid coupling with longitudinal modes, we consider the incompressible limit. This incompressible VE fluid is simulated using a generalized hydrodynamics (GHD) fluid model \cite{singh2014visco,tiwari2015turbulence,dharodi2022kelvin,dharodi2024vortex} under an incompressible limit. The governing equations for incompressible GHD (i-GHD) are detailed in Section \ref{ghd_model}. The VE effects in this model are defined by two coupling parameters: the Maxwell relaxation value $\tau_m$ and the shear viscosity $\eta$. The ratio $\eta/{\tau_m}$ is proportional to the coupling strength of the VE medium. Stated differently, the viscoelastic character of the medium is determined by this ratio. Shear waves in dusty plasmas have been explored analytically~\cite{peeters1987wigner,vladimirov1997vibrational,wang2001longitudinal}, experimentally~\cite{nunomura2000transverse, pramanik2002experimental}, and computationally~\cite{schmidt1997longitudinal}. Kaw et al. \cite{kaw1998low,kaw2001collective} theoretically predicted shear waves using the GHD model. Dharodi et al.~\cite{singh2014visco} computationally confirmed their existence and function under the incompressible limit, demonstrating their ability to suppress gravity-driven instabilities~\cite{das2014collective,dharodi2021numericala,dharodi2021numericalb} and shear driven instabilities~\cite{tiwari2014kelvin,dharodi2022kelvin}, control the evolution of density spiral waves in heterogeneous fluids~\cite{dharodi2020rotating}, influence merging phenomena~\cite{dharodi2024vortex}, and dipole propagation in constant-density dusty plasmas~\cite{dharodi2016sub}.
 \paragraph*{}

\FloatBarrier
In particular, this paper investigates the evolution of two counter-rotating Lamb-Oseen vortices forming a dipole in a two-dimensional SCDP or in a VE fluid, with a focus on the influence of shear waves and density inhomogeneity. The density inhomogeneity is introduced along the vertical y-direction, perpendicular to the moving dipole. Two types of density profiles are considered: (1) a sharp interface separating two distinct density regions, and (2) a gradually varying density profile. For both the density profiles, we have simulated dipole vortices with three different circulation strengths: strong ($\Omega_0=10$), medium ($\Omega_0=5$), and weak ($\Omega_0=3$). For each circulation, we have considered three coupling strengths: mild ($\eta$=2.5, $\tau_m$=20), medium ($\eta$=2.5, $\tau_m$=10), and strong ($\eta$=2.5, $\tau_m$=5). This results in a total of 18 simulation cases, allowing us to explore the impact of dipole vortex circulation strength, medium coupling strength, and background density inhomogeneity on dipole evolution. In VE fluids, a rotating vortex on the low-density side generates faster TS waves than the vortex on the high-density side \cite{dharodi2020rotating}. As a result, we observe that, unlike in the hydrodynamic (HD) case or in a VE medium with constant density~\cite{dharodi2016sub}, the moving dipole structure in an inhomogeneous VE fluid follows a curved upward trajectory. It has been reported that the longevity of a dipole structure is directly proportional to its circulation strength and inversely proportional to the coupling strength of the background medium. A background with high coupling strength supports faster TS waves, which extract energy from the vortex more rapidly, leading to a quicker decay. The circulation strength reflects the amplitude of the dipole structure.

The remaining parts of the manuscript are structured as follows: Section \ref{ghd_model} presents the basic model equations and the implementation of our numerical scheme. To carry out the numerical simulations, the model's equations are first reduced to coupled continuity equations. These reduced equations are then used for the simulation. In Section \ref{Results_discussion}, we discuss the results and observations. Finally, Section \ref{conclusion} provides a summary and concludes the paper.

	\section{The numerical model and simulation methodology} \label{ghd_model}

The generalized hydrodynamic model of a dusty plasma is governed by a set of three coupled equations: the continuity equation (Eq.\eqref{eq:continuity1}), the velocity evolution equation (Eq.\eqref{eq:momentum1}), and Poisson’s equation (Eq.~\eqref{eq:poisson1}).
\begin{equation}\label{eq:continuity1}
  \frac{\partial n_d}{\partial t} + \nabla \cdot \left(n_d\vec{v}_d\right)= 0
  \end{equation}	 
\begin{eqnarray}\label{eq:momentum1}
 &&\left[1 + \tau_m {\frac{d}{dt}}\right]\left[ {{m_d}{n_d}\frac{d\vec{v}_d}{dt}}+{\nabla {p_d}}-{n_d}{Z_d}{e} \nabla \phi_d \right]\\ \nonumber
 &&= \eta \nabla^2 \vec{v}_d+{\left({\zeta+\frac{\eta}{3}}\right)}{\nabla}{\left(\nabla \cdot \vec{v}_d\right)}
\end{eqnarray}
\begin{equation}\label{eq:poisson1}
 \nabla^2 \phi_d  ={-4\pi e}\left(n_i - n_e-{Z_d}{n_d} \right)
\end{equation}
Here, $\zeta$ denotes the bulk viscosity coefficient, and the total time derivative is defined as ${d}/{dt} = \left({\partial}/{\partial t} + \vec{v}d \cdot \nabla \right)$. The dust fluid velocity, electrostatic potential, and number densities of the charged species (electrons, ions, and dust) are represented by $\vec{v}d$, $\phi_d$, and $n_s$$(s = e, i, d)$, respectively. To express Eqs.(\ref{eq:continuity1})–(\ref{eq:poisson1}) in dimensionless form, the following normalization factors are used: the inverse dust plasma frequency $\omega{pd}^{-1} = \left(4\pi (Z_d e)^2 n{d0} / m_d\right)^{-1/2}$ for time, the Debye length $\lambda_d = \left(K_B T_i / 4\pi Z_d n_{d0} e^2\right)^{1/2}$ for length, $\lambda_d \omega_{pd}$ for velocity, and $Z_d e / K_B T_i$ for potential. The dust charge $Z_d$ is assumed to be constant. Here, $m_d$, $T_i$, and $K_B$ denote the dust grain mass, ion temperature, and Boltzmann constant, respectively. The densities $n_s$ ($s = e, i, d$) are normalized using their respective equilibrium values $n_{s0}$.The pressure is determined using the equation of state: $p_d = \mu_d \gamma_d n_d K_B T_d$, where $\mu_d = \left(1 / T_d\right) \left(\partial p_d / \partial n_d\right)_{T_d}$ is the compressibility parameter, and $\gamma_d$ is the adiabatic index. The normalized continuity, momentum, and Poisson equations then become:
\begin{equation}\label{eq:continuity2}
  \frac{\partial \rho_d}{\partial t} + \nabla \cdot
\left(\rho_d\vec{v}_d\right)=0{,}
  \end{equation}
\begin{eqnarray}\label{eq:momentum2a}
 &&\left[1 + {\tau_m}\frac{d} {dt}\right]
  \left[ {{\rho_d}\left(\frac{\partial{\vec{v}_d}}{\partial {t}}+{\vec{v}_d}{\cdot} \nabla{\vec{v}_d}\right)}+{\nabla}{p_d}+{\rho_c}\nabla \phi_{d} \right]\nonumber\\
	&&= \eta \nabla^2 \vec{v}_d+{\left({\zeta+\frac{\eta}{3}}\right)}{\nabla}{\left(\nabla \cdot \vec{v}_d\right)}
  \end{eqnarray}
\begin{equation}\label{eq:poisson2}
 \nabla^2 \phi_d  = \left[n_d + \mu_{e}exp(\sigma_e\phi_d) - \mu_{i}exp(-\phi_d)\right].
\end{equation}
The mass density of the dust fluid is given by 
${\rho_d}= {n_d}{m_d}$. The parameters $\sigma_e = {T_i}/{T_e}$, $\mu_{e} ={n_{e0}}/{Z_{d}n_{d0}}$ and $\mu_{i} ={n_{i0}}/{Z_{d}n_{d0}}$ are defined.  It is reasonable to assume that both electrons and ions follow a Boltzmann distribution, as their inertia is negligible on the slow dust time scales. In the incompressible limit, the normalized continuity and momentum equations can be expressed as:
\begin{equation}\label{eq:continuity}
  \frac{\partial \rho_d}{\partial t} + \nabla \cdot
\left(\rho_d\vec{v}_d\right)=0{,}
  \end{equation}
\begin{eqnarray}\label{eq:momentum2}
	&&\left[1+{\tau_m}\left(\frac{\partial}{\partial{t}}+{\vec{v}_d}\cdot \nabla\right)\right]\nonumber\\
	&& \left[ {{\rho_d}\left(\frac{\partial{\vec{v}_d}}{\partial {t}}+{\vec{v}_d}{\cdot} \nabla{\vec{v}_d}\right)}+{\nabla}{p_d}+{\rho_c}\nabla \phi_{d} \right]\nonumber\\
	&& =\eta \nabla^2\vec{v}_d{,}
\end{eqnarray}
respectively, and the incompressible state is defined by 
\begin{equation}
\label{eq:incompressible}
{\nabla}{\cdot}{\vec{v}_d}=0{.}
\end{equation}
In previous works~\cite{dharodi2014visco,dharodi2016sub}, the procedure for the numerical implementation and validation of these normalized equations has been demonstrated. Here, we consider a constant charge density $\rho_c$, which can be either negative or positive~\cite{shukla2015introduction,chaubey2021positive,dharodi2023ring,chaubey2023controlling}. In the case of constant mass density, i.e., 
$  {\rho_d}(x,y,t)={\rho_{cd}}$
 , Eq.~\eqref{eq:momentum2} becomes:

\begin{equation}\label{eq:momentum3}
\begin{aligned}
& \left[1 + \tau_m \left(\frac{\partial}{\partial t}+\vec{v}_d  \cdot \nabla\right)\right] \\
& \left[{\frac{\partial \vec{v}_d } {\partial t}+\vec{v}_d  \cdot{\nabla}\vec{v}_d } + \frac{\nabla {p_d}}{\rho_{cd}} -\frac{\rho_c}{\rho_{cd}} {\nabla {\phi_d}} \right]  =  {\eta'}\nabla^2 \vec{v}_d{.}
 \end{aligned}
  \end{equation}
where ${\eta'}={{\eta}/{\rho_{cd}}}$.  In the regime, {${\tau_m}{{\partial}/{\partial{t}}} \geq 1$}, taking the curl of Eq.~\eqref{eq:momentum3} and retaining only the linearized terms yields a model equation that supports the propagation of transverse waves with a phase velocity given by:
\begin{equation}\label{eq:TSwave_conf}
	{v_p}=\sqrt{{\eta^{'}}/{{\rho_d}{\tau_m}}}
    \end{equation}
it is proportional to the inverse square root of the medium's density ${\rho_d}$ and the square root of the coupling strength $\eta'/{\tau_m}$. In other words, a medium with higher coupling strength and lower density supports faster TS waves, and vice versa. This relation indicates that the viscosity 
${\eta'}$, in conjunction with the elasticity or relaxation time parameter 
$\tau_m$, plays a significant role in supporting the transverse mode. A detailed derivation of this result can be found in ~\cite{dharodi2014visco,dharodi2016collective,dharodi2020rotating}. Hereafter, we will denote 
$\eta'$  simply as $\eta$, as we have assumed 
$\rho_{cd}$=1 throughout the article.
\subsection*{Simulation methodology}
	\label{num_methodology}
For the numerical simulation, the Eq.~\eqref{eq:momentum3} is reformulated as the following set of two coupled convective equations,
\begin{eqnarray}\label{eq:vort_incomp1}
 {\frac{\partial \vec{v}_d } {\partial t}+\vec{v}_d  \cdot{\nabla}\vec{v}_d } + {\nabla {p_d}} -{\rho_c} {\nabla {\phi_d}} ={\vec \psi}
\end{eqnarray}
	\begin{equation}\label{eq:psi_incomp1}
	\frac{\partial {\vec \psi}} {\partial t}+\vec{v}_d \cdot \nabla{\vec \psi}=
	{\frac{\eta}{\tau_m}}{\nabla^2}{\vec{v}_d }-{\frac{\vec \psi}{\tau_m}}{.}
	\end{equation}
    In our 2D system, the variables vary in  x and y directions. The quantity ${\vec \psi}(x,y)$ is the strain created in the elastic medium by the time-varying velocity fields. In the form of vorticity ($ {\xi_z}={\vec \nabla}{\times}{\vec v_d} $) the above set of becomes
	\begin{equation}\label{eq:cont_incomp3}
	\frac{\partial \rho_d }{\partial t} +  \left(\vec{v}_d\cdot
	\nabla\right)\rho_d= 0{,}
	\end{equation}
	\begin{equation}\label{eq:psi_incomp3}
	\frac{\partial {\vec \psi}} {\partial t}+\left(\vec{v}_d \cdot \vec
	\nabla\right)
	{\vec \psi}={\frac{\eta}{\tau_m}}{\nabla^2}{\vec{v}_d }-{\frac{\vec
			\psi}{\tau_m}}{,}  
	\end{equation}
	\begin{equation}\label{eq:vort_incomp3} 
	\frac{\partial{\xi}_z} {\partial t}+\left(\vec{v}_d \cdot \vec \nabla\right)
	{{\xi}_z}={\frac{\partial}{\partial x}}\left({\frac{\psi_{y}}{\rho_d}}\right)
	-{\frac{\partial}{\partial y}}\left({\frac{\psi_{x}}{\rho_d}}\right){.}   
	\end{equation}
    We employed the LCPFCT method (Boris {\it et al.}~\cite{boris_book}) to evolve the coupled set of Eqs. (\ref{eq:cont_incomp3}),~  (\ref{eq:psi_incomp3}) and (\ref{eq:vort_incomp3}). This method is based on a finite difference scheme associate with the flux-corrected algorithm. The velocity at each time step has been updated by using the Poisson's equation ${\nabla^2}{\vec{v}_d}=-{{\nabla}}{\times}{\vec \xi}$. This Poisson's equation has been solved by using the FISPACK~\cite{swarztrauber1999fishpack}. 
\section{Numerical simulation results}
\label{Results_discussion}
The Lamb-Oseen vortex is a classic solution to the incompressible Navier-Stokes equations in an unbounded domain. A single circular Lamb-Oseen vortex \cite{colagrossi2016particle,zuccoli2024trapped} with circulation ${\Omega_0}$ is defined as:
\begin{equation}
\label{eq:dipole_profile} 
{\omega_{0}(r,t=0)}={\frac{\Omega_0}{2\pi{a^2_0}}}{{e^{-{r^2}/{a^2_{0}}}}}{.}
\end{equation}
where, $r^2={(x-x_0)^2+(y-y_0)^2}$, $a_0$ is the initial vortex radius, and ${\Omega_0}$ is the circulation of the vortex. In this study, the flow is initiated by superposing two axisymmetric, counter-rotating Lamb-Oseen vortices, which have zero net angular momentum, given as
 \begin{eqnarray}
 \label{eq:dipole_profile} 
 {\omega_{0}(r_1,r_2,t=0)}={\frac{\Omega_1}{\pi{a^2_1}}}{{e^{-{r_1^2}/{a^2_{1}}}}}-
  {\frac{\Omega_2}{\pi{a^2_2}}}{{e^{-{r_2^2}/{a^2_{2}}}}}{.}
 \end{eqnarray}
Here $r_1^2={\left(x-x_{01}\right)^2-\left(y-y_{01}\right)^2}$ and $r_2^2={\left(x-x_{02}\right)^2-\left(y+y_{02}\right)^2}$, $a_1$ and $a_2$ are the initial radii of two vorticies, $b_0={y_{02}}-{y_{01}}$ is the separation distance between the vortex cores, and $\Omega_1$ and $\Omega_2$ are the circulations of the two vorticies. In this study, we consider equal circulations for both, i.e., $\Omega_1 = \Omega_2 = \Omega_0$. Figure 1(a) displays the schematic profiles of a Lamb-Oseen vortex pair and the corresponding velocity vector field.  Here, the top vortex rotates counterclockwise, while the bottom one rotates clockwise, resulting in net horizontal motion as a single entity along the x-axis if no dissipation effect is involved.
\begin{figure}[ht]
   \centering    
   \includegraphics[width=1.0\linewidth]{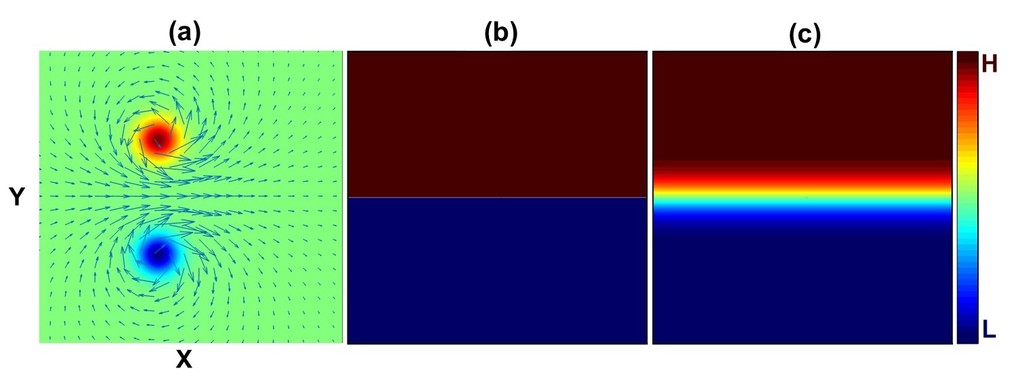}
\captionsetup{justification=raggedright, singlelinecheck=false}
    \caption{Schematic diagrams illustrating (a) a counter rotating Lamb-Oseen vortex pairs along with quiver plot that indicates the velocity associated with the dipolar structure; and density profiles considered initially i.e. at $t=0$ in our simulations  (b) a sharp density interface as well as (c) a gradually changing density gradient along $y-$ axis; The abbreviations 'H' and 'L' written on the side color bar respectively indicate the high and low value of the corresponding physical quantities. }
    \label{fig:figure1}
\end{figure}
We investigate how this dipolar structure propagates along x-axis in background inhomogeneous density along the y-axis, which is provided as 
\begin{eqnarray}
\label{eq:dens_profile} 
{\rho_{0}(y,t=0)}=a_0+b_0tanh(y/\sigma_0)
\end{eqnarray}
Here, $\sigma_0$ is the parameter that controls the sharpness of the density profile. $a_0=1.5$ represents the background density, and $b_0=0.5$ controls the strength of the density variation in the y direction. We consider two main different density profiles: (A) with a sharp density interface for $\sigma_0$=0.05, as illustrated in Fig. \ref{fig:figure1}(b), where the denser fluid occupies the upper region ($\rho_{d02}$, red) and the lighter fluid the lower region ($\rho_{d01}$, blue); and (B) a gradual density gradient for $\sigma$=5.0, as illustrated in Fig. \ref{fig:figure1}(c). Since there is no density stratification in our system against an accelerating force like gravity, this initial condition has nothing to do with the Rayleigh-Taylor instability and buoyancy-driven instability \cite{dharodi2021numericala, dharodi2021numericalb,dharodi2024numerical}. Three circulation strengths—weak ($\Omega_0=3$), medium ($\Omega_0=5$), and strong ( $\Omega_0=10$)—are simulated as cases (1), (2), and (3), respectively. These three different circulation strengths make the dipole move with the axial speeds $v_{dip}$ as fast for $\Omega_0=10$, sluggish for $\Omega_0=5$, and slowest for $\Omega_0=3$. Table~\ref{table:table1} provides an overview of these configurations, which will be explored in the following sections.
\begin{table}
 \begin{tabular}{|l|*{3}{c|}}\hline
\hbox{Case}
&\makebox[11em]{Circulation strength ($\Omega_0$)}
&\makebox[13em]{Dipole propagation speed ($v_{dip}$)}
\\\hline
{(1)} &Strong ($\Omega_0=10$) & {Fast ($v_{dip}^{\Omega_{0}=10}$)}\\\hline
{(2)} &Medium ($\Omega_0=5$)& {Sluggish ($v_{dip}^{\Omega_{0}=5}$)}\\\hline
{(3)} & Mild ($\Omega_0=3$)& {Slowest ($v_{dip}^{\Omega_{0}=3}$)}\\\hline
\end{tabular}
\captionsetup{justification=raggedright, singlelinecheck=false}
\caption{A table listing the cases that have been examined in this article.}
\label{table:table1}
\end{table}

All numerical simulations are performed on a $512{\times}512$ grid in the x and y directions within the simulation domain. To verify the independence of the numerical results from the grid resolution, a grid convergence analysis was performed for each case. The boundaries along the $x-$ axis are considered periodic while they are non-periodic along the $y-$ axis. Throughout the paper, the density vortex is referred as a 'blob'. 
\subsection*{Sharp interface ($\sigma_0=0.05$) :}
\label{Sub:Sharp interface_num}
The profile of the sharp density interface (Fig.~\ref{fig:figure1}(b)) is composed of two incompressible fluids with constant densities, where $\rho_{d01}$=1 for $-12{\pi}\leq y \leq 0$ (bottom half or low-density region) and $\rho_{d02}$=2 for $0 \leq y \leq 12{\pi}$ (upper half or high-density region). The dipole structure is placed at the interface ($({y_{01}},{y_{02}})=(-3a_0,3a_0)$) of these two different densities. A VE fluid allows TS waves to be emitted from a rotating vorticity with velocity, ${v_p}=\sqrt{{\eta}/{{\rho_d}{\tau_m}}}$, which is proportional to the inverse square root of the medium's density ${\rho_d}$ and the square root of the coupling strength of the medium $\eta/{\tau_m}$. In other words, a medium with higher coupling strength and low density supports the faster TS waves and vise-versa. We consider three coupling regimes for each circulation strength (discussed in Table~\ref{table:table1}): mild ($\eta$=2.5, $\tau_m$=20), medium ($\eta$=2.5, $\tau_m$=10), and strong ($\eta$=2.5, $\tau_m$=5). Table~\ref{table:table2} presents the speeds of TS waves in mediums with varying densities and coupling strengths, which are analyzed in the following subsections.
\begin{table}
 \begin{tabular}{|l|*{2}{c|}}\hline
\backslashbox{Coupling strength}{{\hspace{-0.8cm}} Speed of TS ${v_p}$ waves}
&\makebox[6em]{${v_p}$ for $\rho_{d01}$=1}
&\makebox[6em]{${v_p}$ for $\rho_{d02}$=2}
\\\hline
{$\eta$=2.5, $\tau_m$=5} ~~(strong)& {0.71}& {0.5}\\\hline
{$\eta$=2.5, $\tau_m$=10} ~(medium)& {0.5}& {0.35}\\\hline
{$\eta$=2.5, $\tau_m$=20} ~~(mild)& {0.35} & {0.25}\\\hline
{Inviscid fluid} & {0} & {0}\\\hline
\end{tabular}
\captionsetup{justification=raggedright, singlelinecheck=false}
\caption{A table summarizing the TS wave speed for the cases examined in this article.}
\label{table:table2}
\end{table}

\begin{figure}
   \centering    \includegraphics[width=1.0\linewidth]{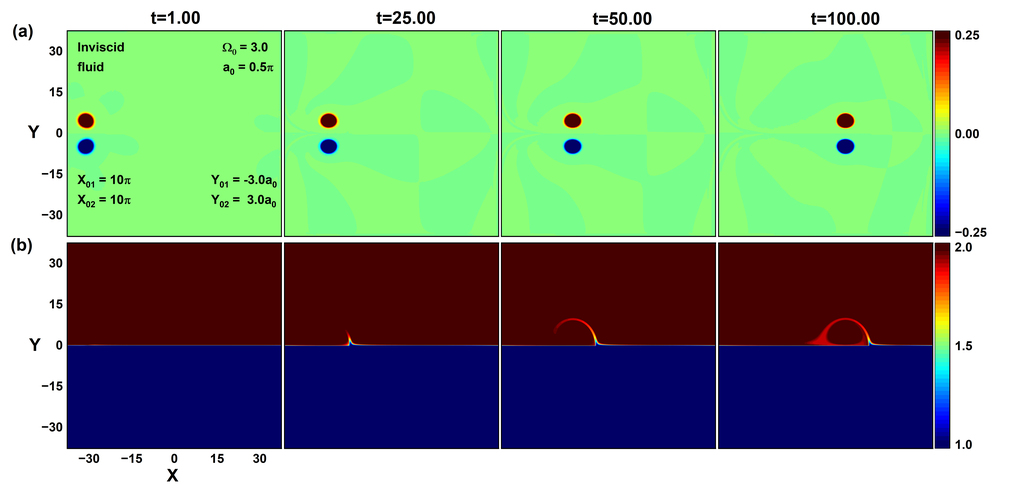}
\captionsetup{justification=raggedright, singlelinecheck=false}
    \caption{Dipole Propagation (\textcolor{blue}{for} $\Omega_0=3$) in an inviscid HD fluid with a sharp background density interface. (a) Time evolution of dipolar structure. The vortices maintain a steady distance from each other and move along the axis at a constant speed. (b) Time evolution of sharp density background profile: Both rotating vortices convect material from the interface, leading to the formation of an expanding envelope of density arms around the moving vortices over time in the inhomogeneous density medium.}
    \label{fig:figure2}
\end{figure}

 First, we investigate the propagation of a dipole in an incompressible, inviscid fluid for all three cases of the circulation strengths. We found that, in the inviscid case, the counter-rotating vortices propagate along their axis as a single stable entity, according to their respective circulation strengths. Moreover, we observe that regardless of the circulation strengths (fast for $\Omega_0=10$, sluggish for $\Omega_0=5$, and slowest for $\Omega_0=3$), the vortices maintain a steady distance from each other and move along the axis at a constant speed.  
So, for inviscid fluids, here we discuss only the case of a vorticity profile that has circulation strength $\Omega_0=3$ and the results are shown in Fig. \ref{fig:figure2}. Figure \ref{fig:figure2}(a) shows the time evolution of the vorticity, and corresponding to this vorticity, the evolution of the background density is shown in Fig. \ref{fig:figure2}(b), where both rotating vortices convect material along the density interface in the propagation direction of the dipole structure. In particular, the top rotating vortex in the upper half (high-density region) facilitates the penetration of the lower half (low-density region) into the upper half, and vice versa, resulting in a growing envelope of density arms around the moving vortices over time.

Next, we analyze the propagation of the dipole in a VE medium of different coupling strength ($\eta/\tau_{m}$). For each circulation strength, the coupling strength  is varied through the elastic component $\tau_m$ keeping the viscosity constant at $\eta= 2.5$ (see Table~\ref{table:table2}).
\subsubsection{Weak circulation: $\Omega_0=3$; $a_1=0.5\pi$, $a_2=0.5\pi$; $and$ $b_0={3\pi}$}
\label{eq_strength10_eq_size_widely}
\begin{figure}
   \centering    \includegraphics[width=1.0\linewidth]{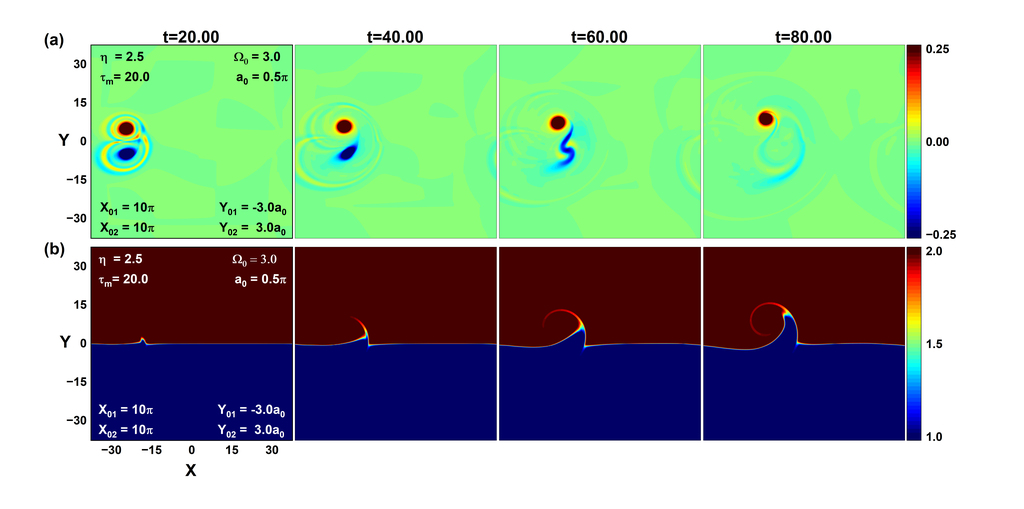}
\captionsetup{justification=raggedright, singlelinecheck=false}
    \caption{Dipole Propagation ($\Omega_0=3$) in VE fluid ($\eta = 2.5$ and $\tau_m = 20$) with a sharp background density interface. (a) Time evolution of dipolar structure: The top vortex in the high-density region emits slower TS waves in comparison to the lower vortex in the low-density region. The faster-emerging waves extract energy from the lower vortex more rapidly, causing it to weaken and decay more quickly than the top vortex. (b) Time evolution of sharp background density profile: where the density perturbations from the interface grow around the stronger vortex in the form of a spiral arm.}
    \label{fig:figure3}
\end{figure}

\paragraph*{}
 First, the value of the circulation strength considered for the dipole is $\Omega_{0}=3$ that is weak in nature i.e. the  dipole moves axially with the slowest speed in comparison to the higher cases of the circulations strengths, $\Omega_{0}=5 \& 10$. In Fig. ~\ref{fig:figure3} we display the results from the simulation for which the coupling parameters are considered as follows, $\eta=2.5$ and $\tau_{m}=20$. Unlike the HD case (Figure~\ref{fig:figure2}), the dipolar structure follows a curved trajectory moving upward. During  propagation, the bottom vortex spiraled around the top vortex and eventually disappeared. As it is well known that a VE fluid supports shear waves. A rotating vortex in the low dense region is expected to generate faster TS waves ($\rho_{01}=1$, $v_p=0.35$) than the top vortexin the high-dense region ($\rho_{02}=1$, $v_p=0.25$) (see Table~\ref{table:table2}). Consequently, shear waves travel a shorter distance in the higher-density region (upper half) than in the lower-density region (lower half) over the same time interval. Figure \ref{fig:figure3}(a) at t = 20 illustrates this effect, showing a larger wavefront of TS waves in the lower half compared to the upper half around their respective vortices. The faster-emerging waves extract energy more rapidly from the lower vortex, causing it to weaken and decay more quickly. The upper vortex, on the other hand, retains its structure with little distortion because energy extraction is slower as the TS waves in the high-density region are not strong enough. This causes the symmetrical dipole structure to transform into an asymmetrical one, in which the top vortex is stronger than the bottom, and to begin moving on a curved path going upward. Between $t=40$ and $t=60$, the weaker vortex tries to spin around the stronger vortex. During this interaction, the weaker vortex stretched and deformed into filament-like structures around the stronger vortex, ultimately leading to its disappearance.  Now, the single stronger vortex remains nearly stationary, rotating about its axis, similar to the case with a constant background density. Figure \ref{fig:figure3}(b) shows the evolution of density corresponding to the evolution of vorticity in Fig. \ref{fig:figure3}(a).
The density perturbations from the interface would be more strongly spiraled around the top-rotating, stronger, and longer-living vortex than around the bottom-rotating vortex. Figure \ref{fig:figure3}(b) shows this effect where the lower region density penetrates through the top density region.

\begin{figure}
   \centering    \includegraphics[width=1.0\linewidth]{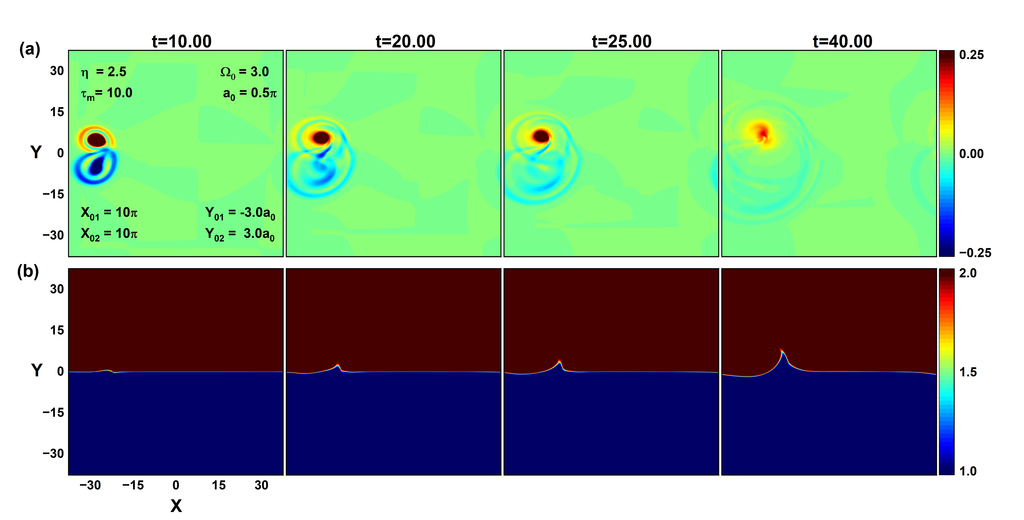}
\captionsetup{justification=raggedright, singlelinecheck=false}
    \caption{Dipole Propagation ($\Omega_0=3$) in VE fluid ($\eta = 2.5$ and $\tau_m = 10$) with a sharp background density interface. (a) Time evolution of dipolar structure: TS waves emerge from vortices faster than in the previous one (Figure \ref{fig:figure3}(a)). Here, the dipolar structure disappears earlier than in \ref{fig:figure3}(a). (b) Time evolution of sharp density profiles: here, since the dipolar structure disappears earlier, that results (Figure \ref{fig:figure3}(b)) in less penetration of the lower region density through the top density region.}
    \label{fig:figure4}
\end{figure}

Figure ~\ref{fig:figure4}  also depicts the evolution of dipole  in a VE fluid but for the increased value of the coupling strength ($\eta = 2.5$ and $\tau_m = 10$). A rotating vortex in the low-dense region ($\rho_{01}=1$, $v_p=0.5$) should generate TS waves faster than that produced in the high-dense region due to top-vortex ($\rho_{02}=1$, $v_p=0.35$). In this case, the TS waves emerge from the vortices travel faster than that in the previous one, which is intended to weaken the strength of the moving dipole. This may be observed by comparing the evolution of vorticity shown in Fig. \ref{fig:figure4}(a) with Fig. \ref{fig:figure3}(a), where the dipolar structure disappears earlier in \ref{fig:figure4}(a) than in \ref{fig:figure3}(a). The evolution of density shown in Figure \ref{fig:figure4}(b) follows the evolution of vorticity in Figure \ref{fig:figure4}(a). Here, since the dipolar structure dissipates earlier, it results in less penetration of the lower-density region into the upper-density region compared to that observed in Fig.~\ref{fig:figure3}(b).

Figure \ref{fig:figure5} shows dipole evolution in a VE fluid for even stronger coupling strength $(\eta = 2.5, \tau_m = 5)$. {In this case, the velocity of the shear waves produced in both high- and low- dense regions due to the respective rotating vortices are respectively $v_{p}=0.71$ and $v_{p}=0.5$. Here, we observe that the  shear waves emitted from the vortex-pair configuration grasp the structure, and tear down them completely without reporting any propagation.  And we can notice its effect on the density evolution (see figure \ref{fig:figure5}(b) ) as well, the density convection due to the rotation of the vortices are almost negligible.
\begin{figure}
   \centering    \includegraphics[width=1.0\linewidth]{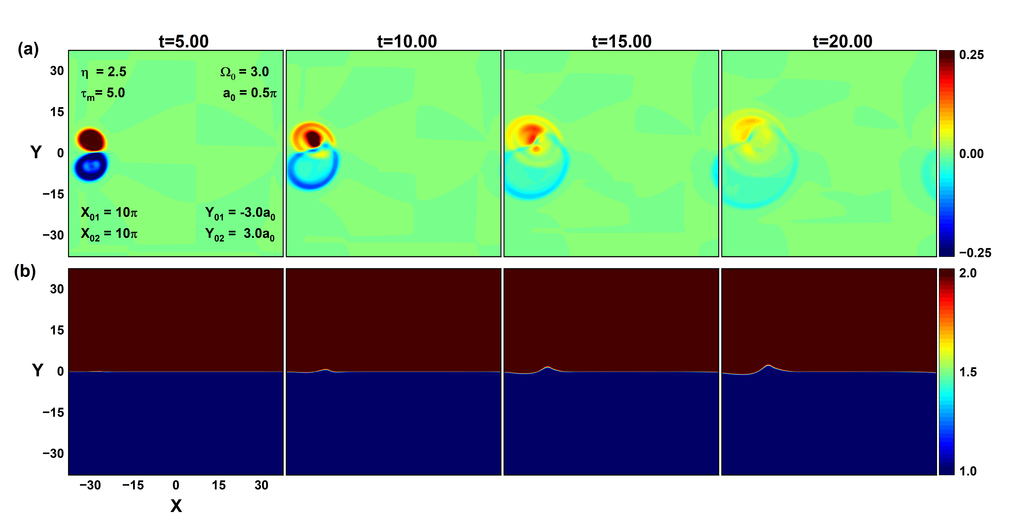}
\captionsetup{justification=raggedright, singlelinecheck=false}
    \caption{Dipole Propagation ($\Omega_0=3$) in VE fluid ($\eta = 2.5$ and $\tau_m = 5$) with a sharp background density interface. (a) Time evolution of dipolar structure: The faster shear waves emitted from the vortex-pair configuration cause the vortices to vanish with almost no propagation. (b) Time evolution of sharp background density profile: The density convection through dipole rotation decreases.}
    \label{fig:figure5}
\end{figure}
\subsubsection{Medium circulation: $\Omega_0=5$; $a_1=0.5\pi$, $a_2=0.5\pi$; $and$ $b_0={3\pi}$}
\label{eq_strength5_eq_size_widely}
\paragraph*{}

For this case, we consider the same initial vortex-separation, same vortex-size and the same set of coupling strength for the background medium, the only physical quantity that differ from the weak case is the circulation strength and its value is  $\Omega_{0}=5.0$. It means that, now, dipole move faster compared to the weak case (see Table~\ref{table:table1}).

Figure~\ref{fig:figure6} shows the evolution of vorticity field  in a VE fluid of coupling parameters, $\eta = 2.5$ and $\tau_m = 20$. This evolution can be directly compared with case (1) of weaker circulation  displayed in Fig.~\ref{fig:figure3}, where the values of all parameters are the same except the circulation strength, which is $\Omega_0 = 3$. Here, the increased circulation strength means the propagating dipole should move faster and survive longer compared to its evolution for weaker case appeared in Figure \ref{fig:figure3}(a). Additionally, the higher strength provides stability to the structure, making it less affected by the emerging shear wave, which results in a less curved trajectory. As can be seen from comparative observations between Figure \ref{fig:figure6}(a) and Figure \ref{fig:figure3}(a). Here, since the dipolar structure survives longer, it results in the density perturbations spiraling around both the vortices, which results in two density blobs, one with high density (dark red) and another with low density (blue), traveling a curved upward path as a single entity over the vorticity dipolar structure, as can be seen in Fig. \ref{fig:figure6}(b).
\begin{figure}
   \centering
\includegraphics[width=1.0\linewidth]{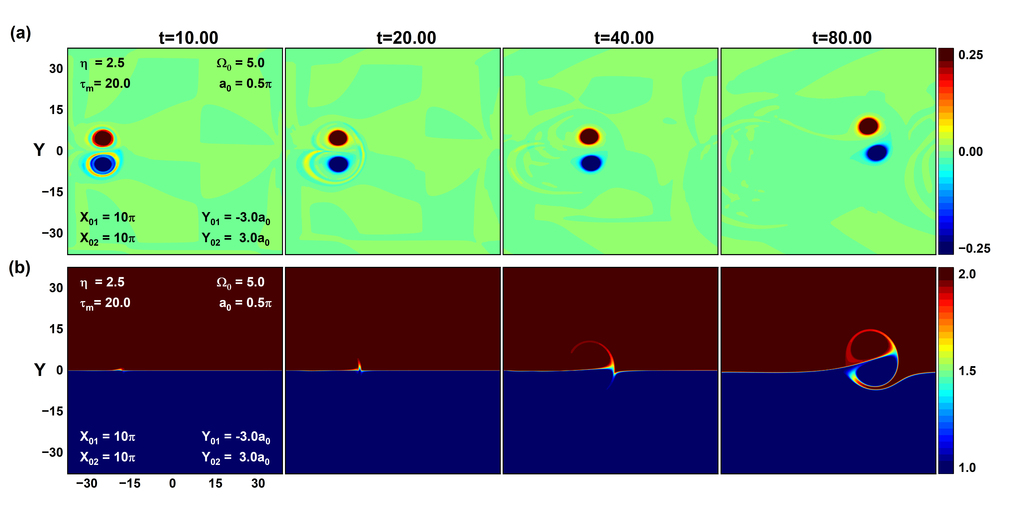}
\captionsetup{justification=raggedright, singlelinecheck=false}
    \caption{Dipole Propagation ($\Omega_0=5$) in VE fluid ($\eta = 2.5$ and $\tau_m = 20$) with a sharp background density interface. (a) Time evolution of dipolar structure: The propagating dipole moves faster and survives longer compared to its evolution in Figure \ref{fig:figure3}(a). (b) Time evolution of sharp density profiles: The density perturbations spiral around both the vortices, which results in two density blobs, one with high density (dark red) and another with low density (blue), traveling a curved path as a single entity.}
    \label{fig:figure6}
\end{figure}


Figures \ref{fig:figure7}(a) and \ref{fig:figure7}(b) show the results for relaxation parameters $\tau_m=10$ and $\tau_m=5$, respectively, and fixed viscosity $\eta=2.5$. The TS waves emerge from vortices more quickly for $\tau_m=5$ than for $\tau_m=10$. In both cases, though, the TS waves appear faster than in the first case discussed in Fig.~\ref{fig:figure6}, which is meant to make the moving dipole less strong. This can be observed by comparing these results: for $\tau_m=10$, the weaker vortex stretched and deformed into filament-like structures around the stronger vortex, and for $\tau_m=5$, the vortices vanished with almost no axial motion while the dipolar structure traveled a curved upward trajectory as a single entity; for $\tau_m=20$, it can be seen from Fig. \ref{fig:figure6}(a). The density evolution be visualized through the density contour snapshots shown in Fig.~\ref{fig:figure8}.
\begin{figure}
   \centering
 \includegraphics[width=1.0\linewidth]{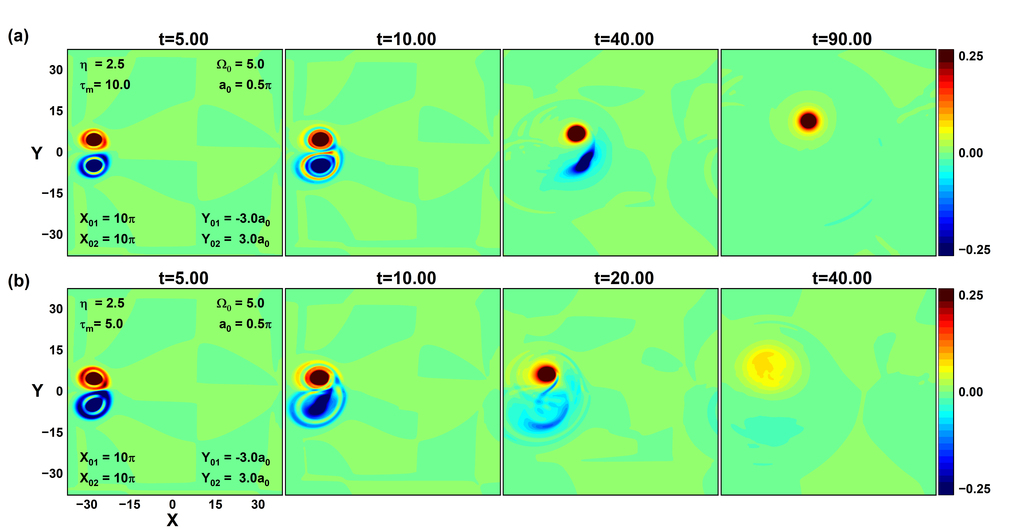}
\captionsetup{justification=raggedright, singlelinecheck=false}
    \caption{Dipole propagation ($\Omega_0=5$) in two VE fluids with sharp background density interfaces: Panel (a) shows a VE fluid with coupling parameters $\eta = 2.5$ and  $\tau_m = 10$, while panel (b) corresponds to a VE fluid with $\eta = 2.5$ and $\tau_m = 5$. It is observed that for $\tau_m=5$, the weaker vortex stretched and deformed into filament-like structures around the stronger vortex, and for $\tau_m=10$, the vortices vanished with almost no axial motion.}
    \label{fig:figure7}
\end{figure}
For $\eta = 2.5$ and $\tau_m = 10$, the lower-density fluid wraps around the upper vortex, penetrating into the high-density region, as the lower vortex vanishes due to stronger TS waves shown in Fig.~\ref{fig:figure7}(b). When $\tau_m$ is further reduced to 5, the penetration of the lower-density fluid is suppressed by even stronger TS waves, causing both vortices to vanish.
   \begin{figure}
   \centering
\includegraphics[width=1.0\linewidth]{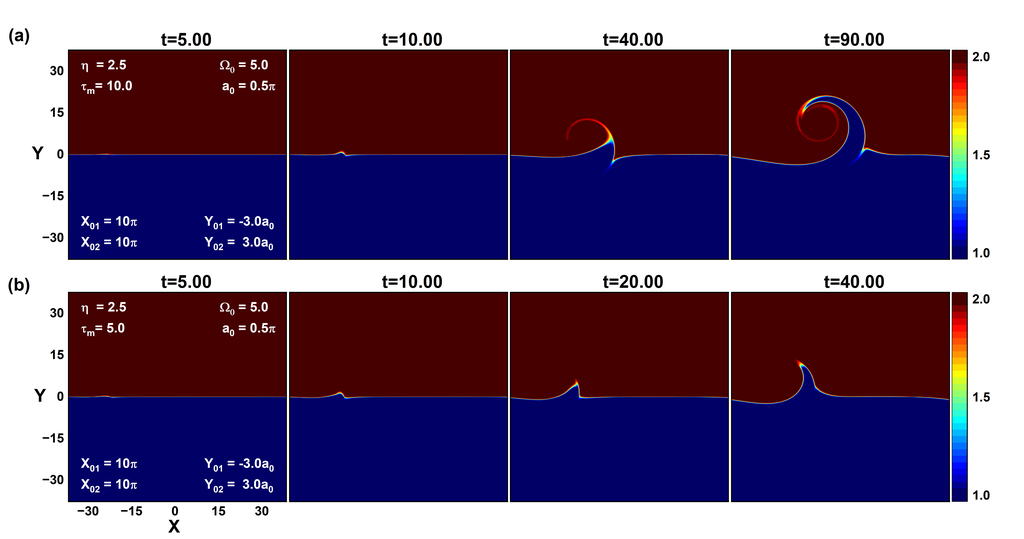}
\captionsetup{justification=raggedright, singlelinecheck=false}
\caption{Time evolution of a sharp background density ($\Omega_0=5$) in two VE fluids, corresponding to the dipole propagation discussed in Fig.~\ref{fig:figure7}: In panel (a) ($\eta = 2.5$ and $\tau_m = 10$), the lower-density fluid wraps around the upper vortex and penetrates into the high-density region, as the lower vortex vanishes due to stronger TS waves. In panel (b) ($\eta = 2.5$ and $\tau_m = 5$), the penetration of the lower-density fluid is further suppressed by even stronger TS waves, leading to the disappearance of both vortices.
}
    \label{fig:figure8}
\end{figure}
\subsubsection{Strong circulation: $\Omega_0=10$; $a_1=0.5\pi$, $a_2=0.5\pi$; $and$, $b_0={3\pi}$}
\label{eq_strength3_eq_size_widely}
\paragraph*{}
The results for the strong circulation with $\Omega_{0}=10$ are displayed in figure \ref{fig:figure9}. The evolution of dipolar structure in VE fluids with fixed viscosity $\eta=2.5$ and relaxation parameters $\tau_{m}=20,10,$ and $5$ is depicted in the top, middle and bottom rows of figure \ref{fig:figure9}, respectively. Increased circulation strength is evident in the dipolar structure's survival even at the medium coupling strength $\tau_{m}=10$ (see Fig. \ref{fig:figure9}(b)), while only stronger vortices survive at $\Omega_{0}=5$ (see Fig. \ref{fig:figure7}(a)) and the entire dipolar structure is engulfed by the TS waves at $\Omega_{0}=3$ (see Fig. \ref{fig:figure4}(a)). The dipolar structure's curved upward trajectory is proportional to the medium's coupling strength, as shown by the comparison of its evolution between Figures  \ref{fig:figure9}(a) and  \ref{fig:figure9}(b).
\begin{figure}
   \centering    \includegraphics[width=1.0\linewidth]{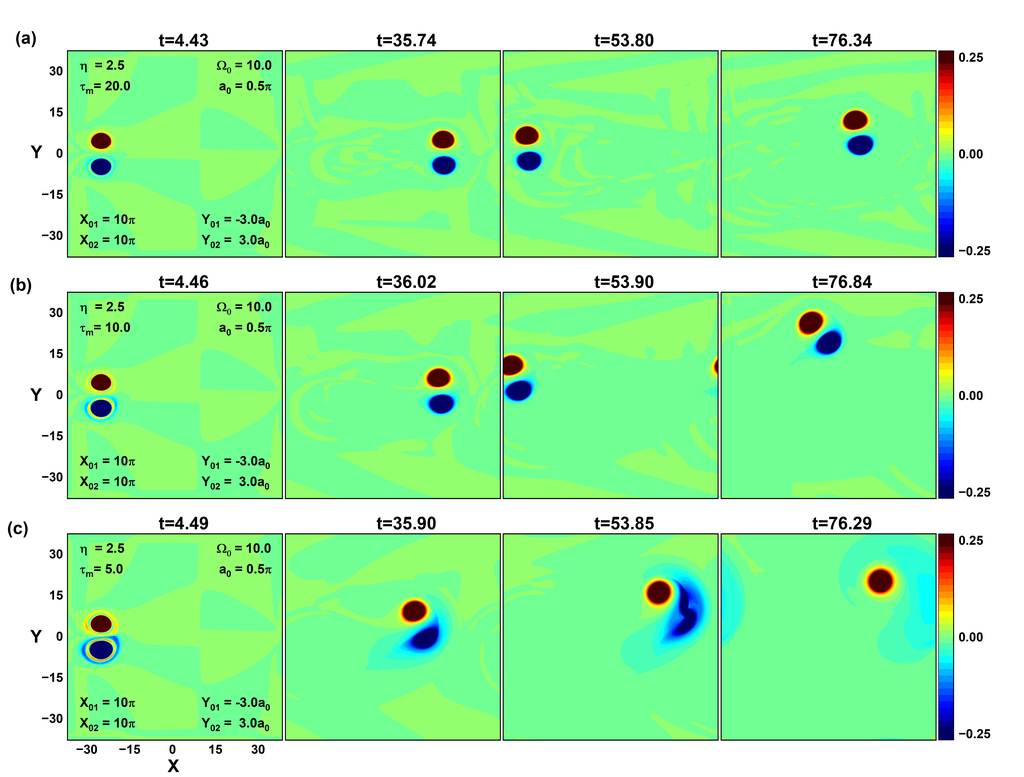}
\captionsetup{justification=raggedright, singlelinecheck=false}
 \caption{Dipole propagation ($\Omega_0=5$) in three VE fluids with a sharp background density interface. In panel (a), the VE fluid has coupling parameters $\eta = 2.5$ and $\tau_m = 20$; in panel (b), $\eta = 2.5$ and $\tau_m = 10$; and in panel (c), $\eta = 2.5$ and $\tau_m = 5$. It can be observed that the upward curvature of the dipolar trajectory increases with the coupling strength of the medium. For $\tau_m = 5$, the bottom vortex spirals around the top vortex and eventually disappears.}
    \label{fig:figure9}
\end{figure}
The corresponding density evolution is given in Fig.~\ref{fig:figure10}. The density blobs become more asymmetrical as the coupling strength grows (from top to bottom row). For $\eta = 2.5$ and $\tau_m = 20$, density perturbations spiral around both vortices, forming high- and low-density blobs that propagate together as a coherent dipole. When $\tau_m$ is reduced to 10, the trajectory of the dipole becomes more curved upward, highlighting the stronger influence of the medium’s coupling strength on its motion. At $\tau_m = 5$, the lower-density fluid wraps around the upper vortex, forming filament-like structures, as the lower vortex vanishes due to the stronger TS waves (see Fig.~\ref{fig:figure9}(c)).
 \begin{figure}
   \centering
\includegraphics[width=1.0\linewidth]{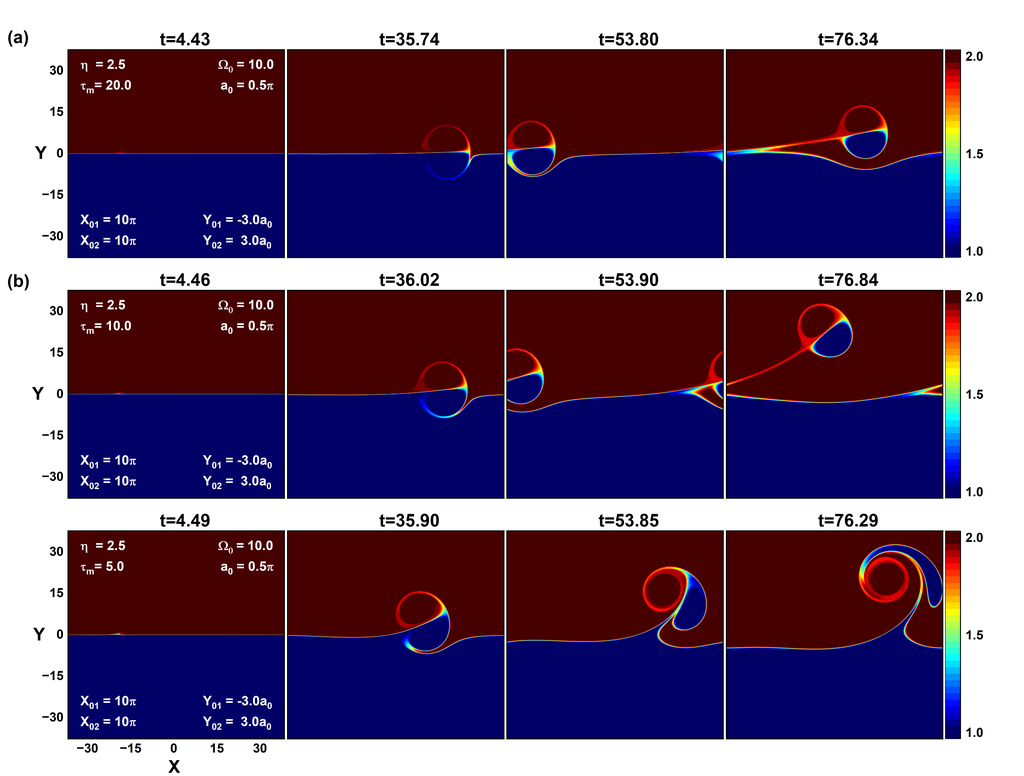}
\captionsetup{justification=raggedright, singlelinecheck=false}
\caption{Time evolution of background sharp density ($\Omega_0=10$) in three VE fluids, corresponding to the dipole evolution discussed in Fig.~\ref{fig:figure9}. As the coupling strength increases (from the top to bottom row), the asymmetry of the density blobs becomes more pronounced.}
    \label{fig:figure10}
\end{figure}

\subsection*{Gradually changing density  ($\sigma_0=5$)}
Further, we perform simulations in which the profile of the dust fluid's density changes progressively, as seen in the Fig.~\ref{fig:figure1}(c).
 
 \subsubsection{Weak circulation: $\Omega_0=3$; $a_1=0.5\pi$, $a_2=0.5\pi$; $and$ $b_0={3\pi}$}

 Figure \ref{fig:figure11}(a) illustrates the time evolution of the dipolar vorticity field in a background medium with coupling parameters  $\eta = 2.5$ and $\tau_m = 20$. Similar to the sharp density case (discussed in Fig.~\ref{fig:figure3}(a)), the dipole structure followed a curved upward trajectory. As it propagated, the bottom vortex spiraled around the top vortex before eventually dissipating. The faster-emerging waves from the lower vortex are causing it to weaken and decay more quickly, while the upper vortex retains its structure with little distortion. This causes the top vortex to be stronger than the bottom and to begin moving on a curved path.  Figure \ref{fig:figure11}(b) shows the evolution of density corresponding to the evolution of vorticity in Fig. \ref{fig:figure11}(a). In Fig. \ref{fig:figure11}(b), the spiral density interface around the vortices is smoother in comparison to the sharp case, as shown in Figure \ref{fig:figure3}(b). Figure \ref{fig:figure11}(b) shows that as the dipole propagates, the density interface around the dipole vortices starts acquiring a mushroom-like shape similar to the gravity-driven instability.  In the beginning, both density blobs are symmetric and travel horizontally, as seen in the snapshots at $t=10$ and $t=20$. However, as the dipole follows a curved upward trajectory—caused by the difference in shear wave speed along the interface—the blobs begin to move along a curved upward path and gradually become asymmetric. This is distinctly observed in snapshots from $t=80$ and $t=100$. 
\begin{figure}
   \centering    \includegraphics[width=1.0\linewidth]{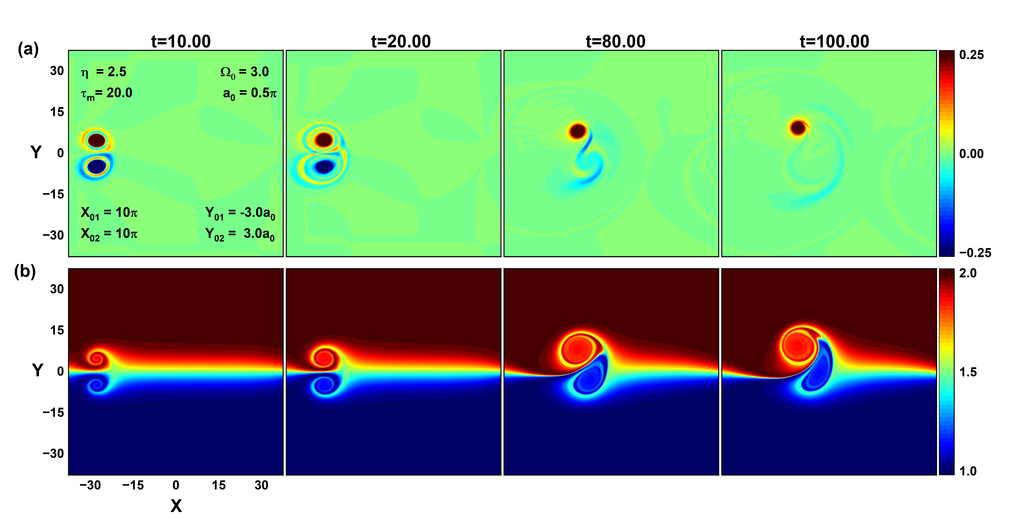}
\captionsetup{justification=raggedright, singlelinecheck=false}
\caption{Dipole Propagation ($\Omega_0=3$) in VE fluid ($\eta = 2.5$ and $\tau_m = 20$) with a gradually changing background density. Top row (a): Time evolution of the dipolar structure — The lower vortex spirals around the upper vortex, gradually weakens, and ultimately vanishes, leading to the distortion of the dipole due to the emerging TS waves. Bottom row (b): Time evolution of gradually changing background density — The density interface evolves into a mushroom-like structure, which progressively deforms over time due to variations in the shear wave speed along the interface.}
    \label{fig:figure11}
\end{figure}

The dipole evolution in a VE fluid for higher coupling strengths ($\eta = 2.5$ and $\tau_m = 10$) is depicted in Figure \ref{fig:figure12}.  In this case, the dipolar structure vanishes earlier (almost before starting the curved trajectory) in Fig. \ref{fig:figure12}(a) than in Fig. \ref{fig:figure11}(a) because of faster emerging TS waves, similar to the sharp density case. The spiral density interface in Fig. \ref{fig:figure12}(b) looks more prominent compared to the sharp density case discussed in Fig. \ref{fig:figure4}(b). Since the dipolar structure vanishes earlier, that results in a mushroom-like density shape that almost travels horizontally but at a later time becomes asymmetric.
\begin{figure}
   \centering    \includegraphics[width=1.0\linewidth]{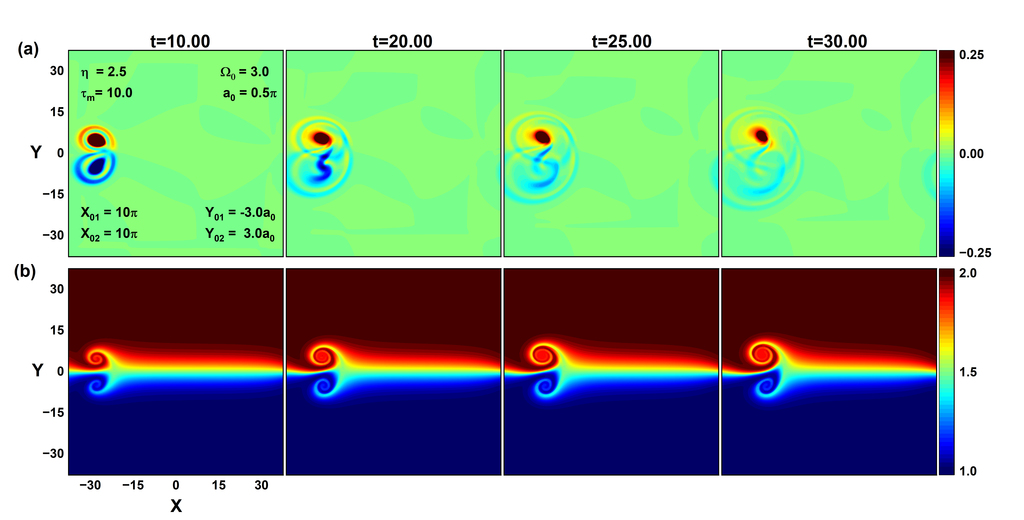}
\captionsetup{justification=raggedright, singlelinecheck=false}
 \caption{Dipole Propagation ($\Omega_0=3$) in VE fluid ($\eta = 2.5$ and $\tau_m = 10$) with a gradually changing background density. Top row (a): Time evolution of the dipolar structure — The dipole deforms earlier compared to the $\tau_m = 20$ case, with the lower vortex losing coherence quickly, leading to a rapid breakdown of the dipolar form. Bottom row (b): Time evolution of gradually changing background density — A mushroom-like density interface emerges.}   
    \label{fig:figure12}
\end{figure}

Dipole evolution in a VE fluid having an even larger coupling strength $(\eta = 2.5, \tau_m = 5)$ is shown in Figure \ref{fig:figure13}. The vortex-pair structure produces quicker shear waves that cause the vortices to disappear with nearly little propagation, much like the sharp density situation described in Figure \ref{fig:figure5}.  Additionally, compared to the two aforementioned scenarios, Figure \ref{fig:figure5}(b) exhibits reduced density convection as a result of vortice rotation.
\begin{figure}
   \centering    \includegraphics[width=1.0\linewidth]{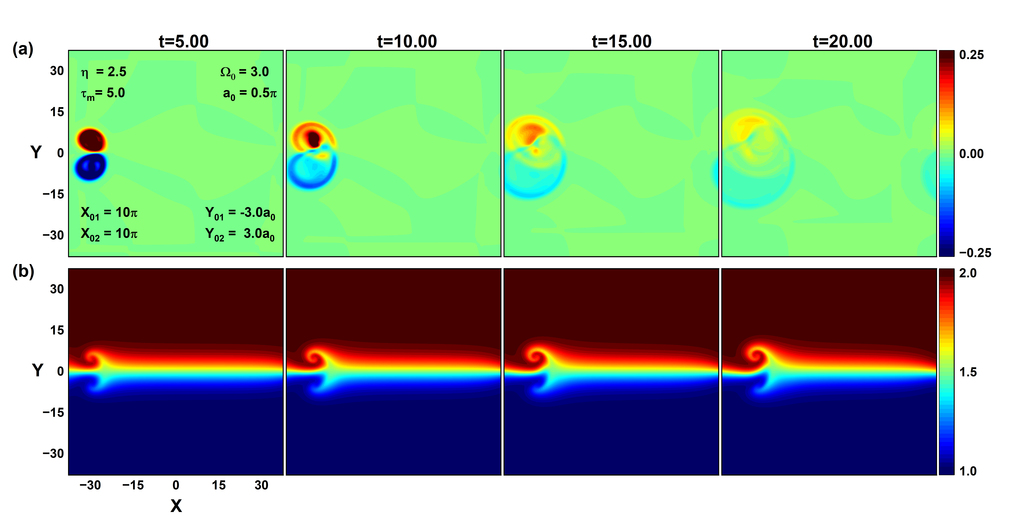}
\captionsetup{justification=raggedright, singlelinecheck=false}
 \caption{Dipole Propagation ($\Omega_0=3$) in VE fluid ($\eta = 2.5$ and $\tau_m = 5$) with a gradually changing background density. Top row (a): Time evolution of the dipolar structure — The dipole exhibits minimal propagation due to rapid emission of shear waves from the vortex-pair configuration. Bottom row (b): Time evolution of gradually changing background density — A smaller mushroom-like structure forms at the interface, indicating reduced density convection resulting from weaker vortex rotation.}  
    \label{fig:figure13}
\end{figure}

\subsubsection{Medium circulation: $\Omega_0=5$; $a_1=0.5\pi$, $a_2=0.5\pi$; $and$ $b_0={3\pi}$}
\label{}
\paragraph*{}
The results are shown with $\Omega_0=5$ strong circulation in Figure \ref{fig:figure14}. The top, middle, and bottom rows of that figure show the evolution of dipolar vorticity for VE fluids with fixed viscosity $\eta = 2.5$ and relaxation parameters $\tau_m$ = 20, 10, and 5, respectively. Compared to the corresponding examples outlined above for $\Omega_0=3$, the dipolar structure remains for a longer period of time, indicating the effect of enhanced circulation strength. As the coupling strength increases (from top to bottom row), the lower vortex disappears more quickly due to stronger transverse shear wave effects, reducing the overall travel distance, since a single vortex does not propagate. In the top row (see Fig. \ref{fig:figure14}(a)), the dipolar structure remains largely intact with minimal decay. In the middle row (see Fig. \ref{fig:figure14}(b)), the dipole structure persists up to t=20, after which the lower vortex vanishes. In the bottom row (see Fig. \ref{fig:figure14}(c)), the dominance of TS effects is evident even at t=10, leading to early distortion and breakup of the dipole structure well before  t=20. The corresponding  density evolution is given in Fig.~\ref{fig:figure15}.
\begin{figure}
   \centering
\includegraphics[width=1.0\linewidth]{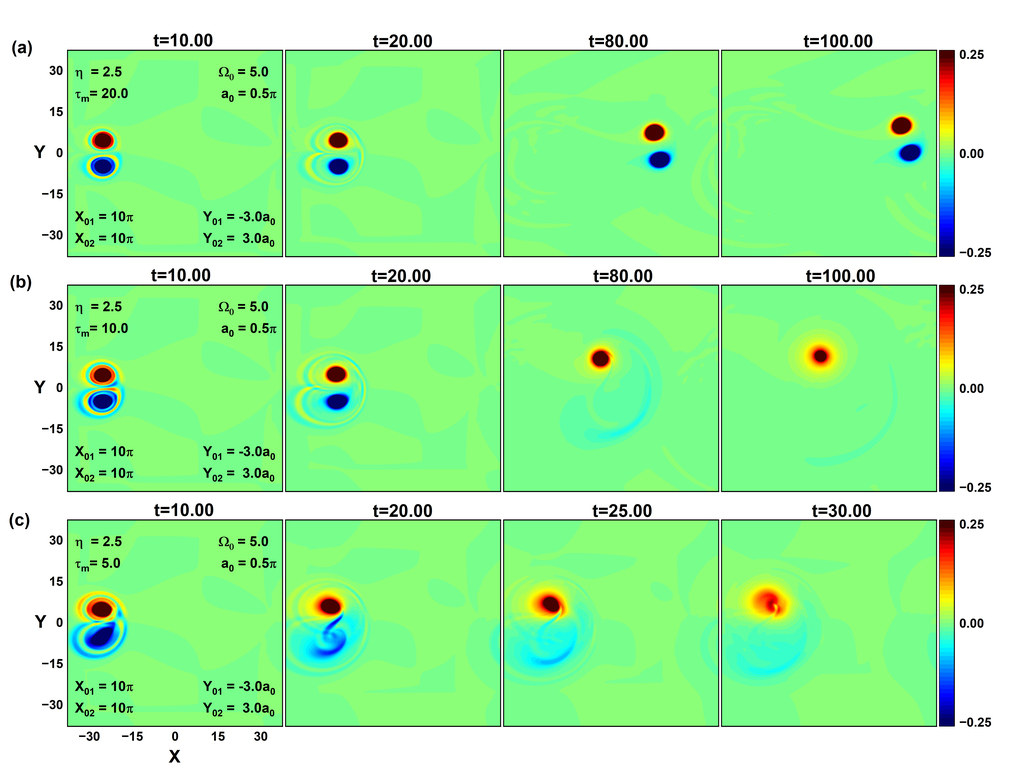}
\captionsetup{justification=raggedright, singlelinecheck=false}
 \caption{Dipole propagation ($\Omega_0=5$) in three VE fluids with a with a gradually changing background density. In panel (a), the VE fluid has coupling parameters $\eta = 2.5$ and $\tau_m = 20$; in panel (b), $\eta = 2.5$ and $\tau_m = 10$; and in panel (c), $\eta = 2.5$ and $\tau_m = 5$. The influence of increasing coupling strength is evident, as the dipolar structure persists for a longer duration from bottom to top row. As the coupling strength increases (i.e., from top to bottom row), the lower vortex vanishes due to stronger shear wave effects, particularly for $\tau_m = 10$ and $\tau_m = 5$. For $\tau_m = 5$, the dipole's travel distance is nearly zero, as a single remaining vortex lacks the ability to sustain convection.}
   
   \label{fig:figure14}
\end{figure}
Similar to the vorticity evolution, the asymmetry in the mushroom-like density structure becomes more pronounced as the coupling strength increases (from top to bottom row). In the bottom row (see Fig.~\ref{fig:figure15}(c)), corresponding to the strongest coupling or strongest TS waves, the density structure begins to deform noticeably by t=20. In the middle row (see Fig.~\ref{fig:figure15}(b)), asymmetry appears around t=80, while in the top row (see Fig.~\ref{fig:figure15}(a)), the dipole structure remains largely symmetric even at 
t=100 and beyond.
  \begin{figure}
   \centering
    \includegraphics[width=1.0\linewidth]{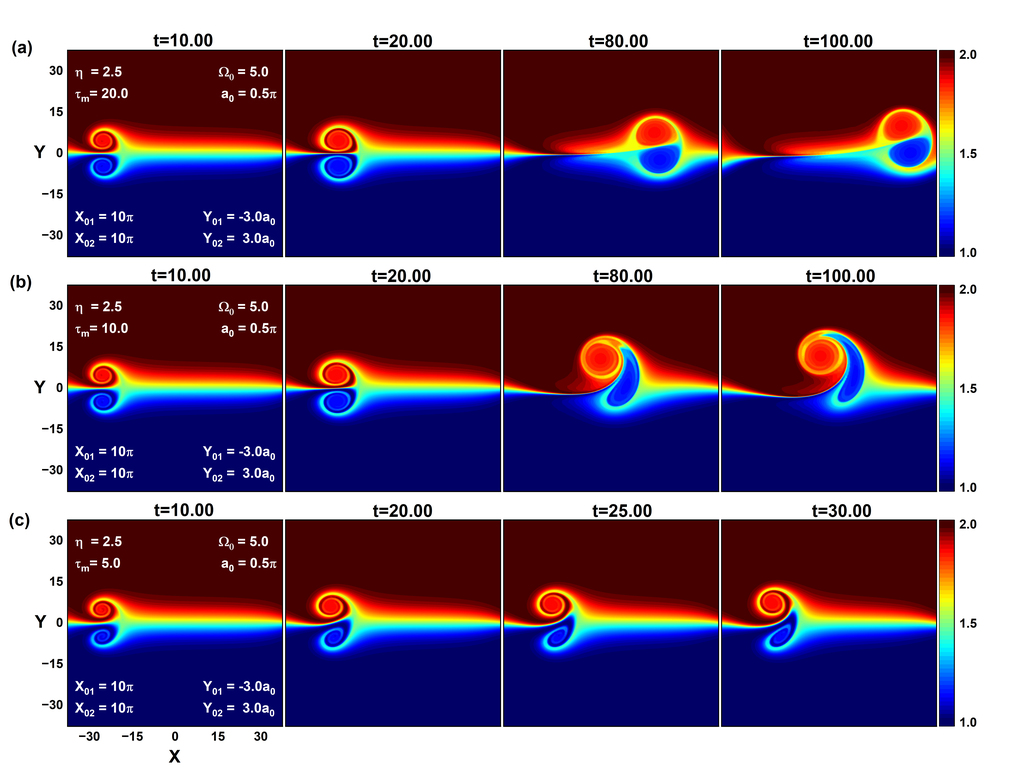}
\captionsetup{justification=raggedright, singlelinecheck=false}
\caption{Time evolution of the smooth density profile ($\Omega_0=10$) in three VE fluids, corresponding to the dipole evolution discussed in Fig.~\ref{fig:figure14}. It can be noticed that the asymmetry in the
mushroom-like density structure becomes more pronounced
as the coupling strength increases (from top to bottom row).}
    \label{fig:figure15}
\end{figure}

\subsubsection{Strong circulation: $\Omega_0=10$; $a_1=0.5\pi$, $a_2=0.5\pi$; $and$, $b_0={3\pi}$}
\label{}
Figure \ref{fig:figure16} displays the results with $\Omega_0=10$ strong circulation. The evolution of dipolar vorticity for VE fluids with fixed viscosity $\eta = 2.5$ and relaxation parameters $\tau_m$ = 20, 10, and 5 is depicted in the top, middle, and bottom rows of that figure, respectively.  One obvious effect of stronger circulation is the longer survival of vortex/vortices in comparison to the two cases discussed above. For both the top (see Fig.~\ref{fig:figure16}(a)) and middle (see Fig.~\ref{fig:figure16}(b)) rows, due to periodic boundary conditions along the x-direction, the dipoles re-enter the domain between t=17.27 and t=63.15. The effect of shear wave strength is evident in the reduced travel distance of the dipole structures.  A comparison of snapshots at $t=63.15$ and $t=77$ shows that the structure in the middle row ($\tau_m=10$) travels a shorter distance and follows a more curved trajectory compared to the top row ($\tau_m=20$), due to stronger shear wave effects. For $\tau_m=5$ (see Fig.~\ref{fig:figure16}(c)), these shear effects are even stronger, leading to a further reduction in travel distance compared to both $\tau_m=10$ and $\tau_m=20$. As a result, the lower vortex rapidly loses its identity, and eventually only the upper vortex survives.
\begin{figure}
   \centering
   \includegraphics[width=1.0\linewidth]{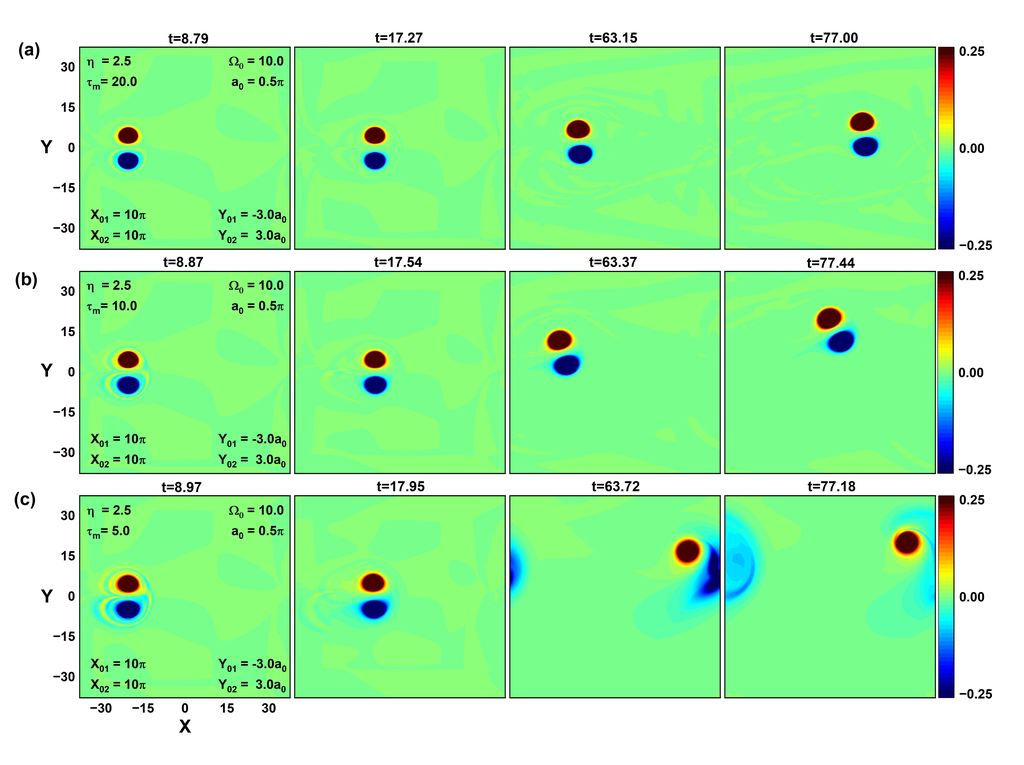}
\captionsetup{justification=raggedright, singlelinecheck=false}
\caption{Dipole propagation ($\Omega_0=10$) in three VE fluids with a with a gradually changing background density. In panel (a), the VE fluid has coupling parameters $\eta = 2.5$ and $\tau_m = 20$; in panel (b), $\eta = 2.5$ and $\tau_m = 10$; and in panel (c), $\eta = 2.5$ and $\tau_m = 5$. The effect of stronger circulation is evident in the prolonged survival of the vortex/vortices compared to the two previously discussed cases. The influence of shear wave strength manifests as a reduction in the dipole’s travel distance and an increase in the curvature of its trajectory.}   
   \label{fig:figure16}
\end{figure}
The corresponding  density evolution is given in Fig.~\ref{fig:figure17}. As the coupling strength increases (from top to bottom row), the asymmetry in the mushroom-like density shape increases. Furthermore, the trajectory of the mushroom-like density for $\tau_m=10$ (see Fig.~\ref{fig:figure17}(b)) is more curved than that for $\tau_m=20$ (see Fig.~\ref{fig:figure17}(a)). For $\tau_m=5$ (see Fig.~\ref{fig:figure17}(c)), the mushroom-like density structure becomes distorted due to the dominance of shear wave emission over the circulation strength of the vortices.
  \begin{figure}
   \centering
\includegraphics[width=1.0\linewidth]{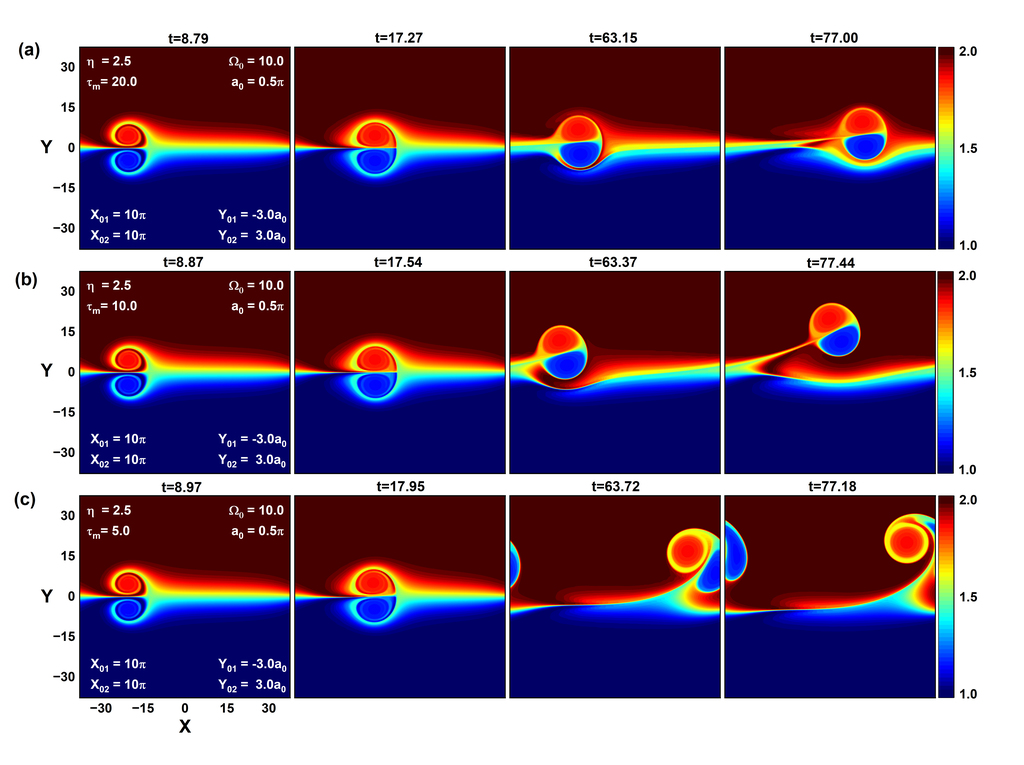}
\captionsetup{justification=raggedright, singlelinecheck=false}
\caption{Time evolution of the smooth density profile ($\Omega_0=10$) in three VE fluids, corresponding to the dipole evolution discussed in Fig.~\ref{fig:figure16}. As the coupling strength increases (from top to bottom row),
the asymmetry in the mushroom-like density shape increases.}
    \label{fig:figure17}
\end{figure}
\section{Conclusion } \label{conclusion}

A propagating dipole structure plays a crucial role in enhancing advection and material transport as it traverses a medium. Such structures are commonly found in a wide range of systems, including hydrodynamic fluids, geophysical flows (such as ocean currents and atmospheric dynamics), plasmas (both laboratory and space plasmas), and astrophysical environments. Their ability to transport mass, momentum, and energy efficiently makes them fundamental to processes like mixing, turbulence generation, and the formation of large-scale coherent structures across different physical settings.

This paper investigates the evolution of two counter-rotating Lamb-Oseen vortices forming a dipole in a two-dimensional SCDP or VE fluid, with a focus on the influence of shear waves and background density inhomogeneity. In SCDPs, modeled as VE fluids, long-lived coherent structures are sustained, significantly affecting instabilities and wave generation. Using the two-dimensional incompressible generalized hydrodynamic model, this work quantitatively examines how density inhomogeneity impacts the dynamics of moving counter-rotating Lamb-Oseen vortices that emit transverse shear waves in incompressible SCDPs. With the dipole positioned at the interface, we examine the dipole's propagation along the x-axis with y-axis density fluctuation, taking into account both sharp and gradual density profiles. The strength of the shear wave emission is determined by the inverse square root of the density of the medium (${\rho_d}$) and the square root of the coupling strength of the medium ($\eta/{\tau_m}$). Therefore, how the interaction between density inhomogeneity and shear wave strength impacts the moving dipole in VE fluids is the primary topic of discussion in this paper.  The following are some key findings. 

  In an inviscid fluid, counter-rotating vortices travel along their axis (horizontal along the x-axis) as a stable unit, spiraling an envelope of density interface around them. In VE fluids, a rotating vortex on the low-density side generate faster TS waves than the vortex on the high-density side (see Table~\ref{table:table2}). The faster-emerging waves extract energy from the lower side vortex more rapidly, causing it to weaken and decay more quickly. This results in the dipole structure following a curved trajectory toward the high-density side.

In the case of a sharp interface ($\sigma_0 = 0.05$), weak circulation in a VE fluid with mild coupling strength ($\eta = 2.5$, $\tau_m = 20$) leads to the decay of the vortex on the low-density side, which spirals around the top vortex at a slower rate compared to the medium coupling case ($\tau_m = 10$). For strong coupling ($\tau_m = 5$), both vortices vanish with minimal axial motion. As circulation increases, the dipole propagates faster and survives longer, with the lower vortex persisting longer and the curved trajectory becoming more pronounced, especially at medium coupling strength. 

For gradually increasing density ($\sigma_0 = 5$), the general effects of circulation and coupling strength are similar to those in the sharp interface case, but notable differences arise in the morphology of the resulting structures. The spiral density interface around the vortices appears smoother compared to the sharp interface case. As the dipole propagates, the density interface around the vortices gradually forms a mushroom-like shape, reminiscent of gravity-driven instabilities. Initially, the density blobs on both sides of the dipole are symmetric and travel horizontally. However, due to the difference in shear wave speeds along the interface, the dipole follows a curved path, causing the density blobs to shift upward and become asymmetric.

These results imply that the propagation of dipole structures can be controlled by varying the medium's coupling strength, which enables control over transport processes like turbulence, diffusion, and mixing. In this study, dipole vortices with identical circulation strengths were considered. Investigating asymmetric dipole propagation in strongly coupled dusty plasmas or other VE fluids would be an interesting direction for future research. Nevertheless, the findings presented here are broadly applicable to any VE medium and are not restricted to dusty plasmas.

\bibliography{ref_dipole}%

\end{document}